\pdfoutput=0
\documentclass[tightenlines,aps,preprint,
showpacs,nofootinbib]{revtex4}

\usepackage{psfig,float}
\usepackage{amsmath,amsfonts,graphicx,bm}

\newcommand{\be}{\begin{equation}}
\newcommand{\ee}{\end{equation}}
\newcommand{\ba}{\begin{eqnarray}}
\newcommand{\ea}{\end{eqnarray}}

\begin{document}


\begin{flushright}
\vbox{
\begin{tabular}{l}
UH-511-1111-07
\end{tabular}
}
\end{flushright}

\title{
NLO QCD corrections to the production of  $t \bar t Z$
in gluon fusion
}

\author{Achilleas Lazopoulos and Kirill Melnikov 
        \thanks{e-mail: kirill@phys.hawaii.edu \\
achilles@phys.hawaii.edu}}
\affiliation{Department of Physics and Astronomy,
          University of Hawaii,\\ 2505 Correa Rd. Honolulu, HI 96822}  
\author{Frank Petriello\thanks{frankjp@phys.wisc.edu}}
\affiliation{
Department of Physics, University of Wisconsin, Madison, WI  53706
\vspace*{1cm}
} 

\begin{abstract}
We compute the ${\cal O}(\alpha_s)$ QCD corrections to  the 
partonic process $gg \to t \bar{t} Z$ at the LHC.  This partonic channel is the dominant component of the scattering process $pp \to t\bar{t}Z$,
which will be important for measuring the $t \bar t Z$ electroweak couplings. The ${\cal O}(\alpha_s)$ corrections
increase the total cross section by up to $75\%$ for reasonable choices of the renormalization and factorization scales.  
Inclusion of these contributions descreases the residual scale dependence of the cross section 
coming from uncalculated higher order terms to $\pm 5\%$.
\end{abstract}

\maketitle


\section{Introduction}

The discovery of the top quark in experiments at the Fermilab 
Tevatron established the validity of the Standard Model as a quantum 
field theory based on the gauge group ${\rm SU}(3) \times {\rm SU}(2) 
\times {\rm U}(1)$ 
with three generations of quarks~\cite{topd}.  Subsequently,
the production cross section for $t \bar t$ pairs
in proton anti-proton collisions was  measured with good precision and 
the value of the top quark mass was established to nearly the percent level~\cite{Pleier:2007dg}. 
Very recently, the first observation of single top production was reported~\cite{stop}.

While the above studies are  important milestones in top quark physics, 
 these measurements probe only a limited subset of top quark properties. 
The process $p \bar p \to t \bar t$ at the Tevatron  is 
sensitive to the top quark mass and color charge.  The decay $t \to Wb$ is 
marginally sensitive to the $tWb$ coupling.  Other important properties of the top quark, 
such as its couplings to the $Z$-boson and the photon and its Yukawa coupling, cannot currently 
be directly accessed experimentally.  These couplings are sensitive probes of new physics 
effects; in particular, the $Zt\bar{t}$ coupling is affected by tree-level mixing with 
additional $Z^{'}$ gauge bosons and vector-like fermions.

While the best place to perform detailed studies of the top quark is the 
International Linear Collider,  useful information about 
the top quark  couplings to electroweak gauge bosons and the Higgs boson
can be obtained from the LHC where the processes $pp \to t \bar t H$, 
$pp \to t \bar t Z$ and $pp \to t \bar t \gamma$ can be observed. However, 
in order to use those observations, 
accurate theoretical predictions for the signal and 
background events are  required.  Since the cross sections for all 
signal processes scale as $\sigma_{\rm LO} \sim \alpha_s^2$ at leading order in the QCD perturbative expansion, 
there is significant uncertainty in the normalization of the 
leading order cross section.  To remove this theoretical error, a computation 
of the cross section through next-to-leading order (NLO) in $\alpha_s$ is required. 

Recently, the possibility of measuring the 
top quark couplings to the photon and the $Z$-boson at the LHC was studied in detail~\cite{baur1,baur2}.  
The study in Ref.~\cite{baur1} required that the $Z$-boson decays leptonically, and allowed for 
either zero or one leptonic $W$ boson decay.  Both channels have a reasonable signal to background 
ratio.  It was pointed out in Ref.~\cite{baur2} that further improvements can be achieved by considering 
the decay $Z \to \nu \bar \nu$.  The background processes to $pp \to t \bar t Z$ and $pp \to t \bar t \gamma$ 
scale with a high power of $\alpha_s$, $\sigma_{\rm bckg} \sim \alpha_s^n$, $n \ge 2$.  
Consequently, theoretical predictions for these processes are quite uncertain.
Nevertheless, as pointed out in Refs.~\cite{baur1,baur2}, it may be possible to 
experimentally study them in signal-free regions
to control the background normalization.
In principle, this technique allows the 
extraction of the $t\bar t Z$ and $t \bar t \gamma $ couplings 
from the $t \bar t Z(\gamma)$ 
cross section measurements, if not for the normalization uncertainty in the signal cross sections themselves
associated with unknown QCD corrections. It was shown in Ref.~\cite{baur1} that this uncertainty
is a limiting factor in extracting anomalous $t \bar{t} Z$ couplings in the leptonic $Z$ decay channels.

In this paper we take the first step towards computing the NLO QCD corrections 
to  $t \bar t Z$ production at the LHC by considering the 
$gg \to t \bar t Z$ partonic subprocess. We focus on this reaction for two reasons.
First, gluon collisions at the LHC are important because of the large gluon 
luminosity. This can be seen already 
at leading order where the $gg \to t \bar t Z$ subprocess 
 gives approximately $60 \%$ of the full result.
Second,  the computation of NLO QCD corrections to the $gg \to t \bar t Z$ subprocess is 
technically more  involved than the corresponding computation for  
quark anti-quark annihilation channels.  
It can also be  expected that NLO QCD corrections to $gg \to t \bar t Z$ are larger than corrections to 
$q \bar q \to t \bar t Z$, since the color charge of gluons is larger than 
that of quarks. For these reasons we believe  that the computation presented 
in this paper gives a good idea about the importance of ${\cal O}(\alpha_s)$ 
corrections  to  the full $pp \to t \bar t Z$ process.
However,  a detailed study of how 
NLO QCD influences  the extraction of the $t \bar t Z$ coupling 
requires the inclusion 
of the $q \bar q$ and $qg$ channels.  We plan to extend our computation to incorporate 
the quark-initiated production processes in the near future.

Our paper is organized as follows. In the next Section we 
introduce our notation and briefly discuss the calculation  of the leading 
order cross section. Section III describes the computation of the ${\cal O}(\alpha_s)$ corrections.
In Section~IV, numerical results are presented and discussed. 
We conclude in Section~V. 

\section{Notation and the leading order cross section}

We consider the process $p(P_1) + p(P_2) \to t \bar t Z$. The 
factorization theorems for hard scattering processes in QCD
allow us to write 
\be
{\rm d} \sigma = \sum_{ij} \int \limits_{0}^{1} {\rm d}x_1 {\rm d}x_2
f^{1}_i(x_1) f^{2}_j (x_2) {\rm d} \sigma_{ij \to t \bar t Z}(x_1 x_2 S),
\label{eq1}
\ee
where $f_{i,j}^{(1,2)}(x_{1,2})$ are the parton densities that give 
the probability to find parton $i(j)$ in the proton $1(2)$ with momentum 
$p_{i(j)} = x_{1(2)} P_{1(2)}$. The center-of-mass energy squared of the 
proton-proton collision is introduced in Eq.~(\ref{eq1}), $S = 2P_1\cdot P_2$.  
In  this paper, we only consider gluon collisions at the LHC, so
we set  $i = j = {\rm gluon}$ in Eq.~(\ref{eq1}) and use 
$\sqrt{S} = 14~{\rm TeV}$.

\begin{figure}[htb]
\includegraphics[angle=0,width=4in,clip=]{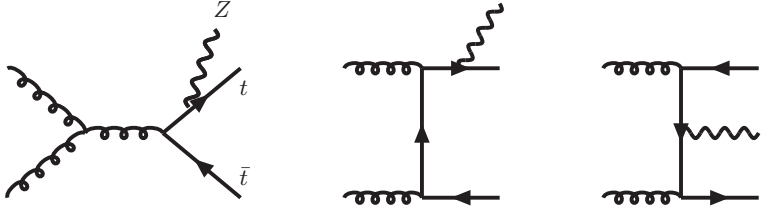}
\caption{Sample diagrams for $gg \to t \bar t Z$ appearing at leading 
order.} 
\end{figure}

At leading order in the $\alpha_s$ expansion, there are 
eight Feynman diagrams that contribute to $gg \to t \bar t Z$; 
some examples  are shown in Fig.1. The computation of the leading order 
cross section is straightforward. We use QGRAF~\cite{qgraf}  
to generate the relevant 
Feynman diagrams and then MAPLE and FORM \cite{form} 
to manipulate this output.
Throughout the paper we set $m_t = 170.9~{\rm GeV}$, $m_Z = 91.18~{\rm GeV}$, and
$m_W = 80.45~{\rm GeV}$.  For the coupling of the $Z$-boson to quarks, 
we employ 
\be
Zqq:~~~i\sqrt \frac{8 m_W^2 G_F}{\sqrt{2} \cos^2 \theta_W} \left ( g^q_v + g^q_a \gamma_5 \right ),
\ee
where $\displaystyle g^q_v = \frac{T_3^q}{2} - Q_q \sin^2 \theta_W,~
g^q_a = -\frac{T_3^q}{2}$,
$\sin^2 \theta_W = 1-m_W^2/m_Z^2 = 0.2215$ is the sine squared 
of the electroweak mixing angle, $T_3^q$ is the weak isospin of the 
quark $q$, $Q_q$ is the electric charge of the quark $q$ in units 
of the proton charge and $G_F$ is the Fermi constant.  Numerical results 
for the leading order cross section are reported in Section IV.

\section{Next-to-leading order computation}

We now discuss the computation of the next-to-leading order corrections 
to the partonic process $gg \to t \bar t Z$.  Three distinct contributions should 
be considered: virtual corrections to the leading order process 
$gg \to t \bar t Z$, the real emission corrections $gg \to t \bar t Z + g$, 
and the renormalization of the leading order cross section.

We first consider the computation of  the virtual corrections 
to $ gg \to t \bar t Z$. There are 162 diagrams that contribute at 
next-to-leading order.  Some examples of these one-loop diagrams are shown 
in Fig. 2.  These diagrams fall into two categories:

\begin{itemize}

\item diagrams in which the $Z$-boson couples to the top quark, such as the first two 
  shown in Fig. 2;

\item diagrams in which the $Z$-boson couples to an internal quark propagator, such as the final 
one in Fig. 2.

\end{itemize}

\noindent
The first class contains contributions proportional to the coupling constant combinations 
$(g^t_v)^2$ and $(g^t_a)^2$.  Diagrams in this class can be computed with standard $d$-dimensional 
Dirac algebra, using the anti-commutation prescription for $\gamma_5$.  The second class contains contributions proportional 
to $g^t_v g^q_v$ and $g^t_a g^q_a$, where $q$ denotes any of the six quarks.  The diagrams in this class contain 
products of two Dirac traces with single $\gamma_5$ terms, leading to contributions containing a product of two Levi-Civita 
tensors.  Although this second set of diagrams is finite, we use the $d$-dimensional prescription for $\gamma_5$ 
described in Ref.~\cite{Larin:1993tq} for calculational convenience in the intermediate steps.

\begin{figure}[htb]
\includegraphics[angle=0,width=4in,clip=]{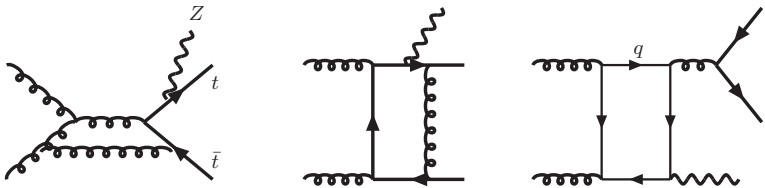}
\caption{Sample diagrams for the ${\cal O}(\alpha_s)$ virtual 
corrections to 
$gg \to t \bar t Z$.} 
\end{figure}

To compute the virtual corrections, we employ the method described in Ref.~\cite{lmp}.  We summarize here the salient features of this technique. 
Further details can be found in Ref.~\cite{lmp}.
We compute the square of the scattering amplitude, summing over the fermion 
spins and polarizations of gluons and the $Z$-boson. For each  
one-loop diagram interfered with the full Born amplitude, we perform the integration  
over the loop momentum and arrive at an integral over 
Feynman parameters.  The integration over Feynman parameters cannot be computed 
numerically because it diverges. While the ultraviolet 
divergencies factorize 
after the integration over the loop momentum and 
therefore have no effect on the integration  over Feynman parameters, 
the infrared and collinear singularities appear when certain Feynman parameters
approach boundary values.  To have a representation of the integrals 
 suitable for numerical evaluation, we must extract the infrared and collinear singularities.  We accomplish this 
using the method of sector decomposition \cite{bh}. However, even after 
these divergencies are extracted, it is not possible to perform the Feynman parameter integrations numerically because there are singularities 
inside the integration region coming from internal loop thresholds.  These singularities are avoided by deforming 
the integration contour into the complex plane following 
the suggestion in Ref.~\cite{soper}. Once  
infrared and collinear singularities are extracted and the integration 
contour is deformed, 
we obtain representations of the Feynman parametric 
integrals suitable for numerical integration.  This integration 
is performed using the adaptive Monte Carlo integration 
algorithm VEGAS \cite{vegas} as implemented in the CUBA library \cite{cuba}.

We briefly comment here on the stability of the numerical integration. 
In principle, after extracting infra-red and collinear singularities 
and deforming the contour into the complex plane, the 
integration should be numerically stable.  For most phase space 
points that we consider, we find this to be the case. However, there are 
situations where straightforward  application of the algorithm described in 
\cite{lmp} leads to numerical instabilities. These instabilities typically 
occur when the sector decomposition procedure ``overdoes'' the extraction of 
singularities, splitting the integral into too many sectors which each contain unphysical singularities.  
Problems of these type occur typically 
in the evaluation of five-point functions.   We handle these problematic 
cases by choosing a more sophisticated parameterization of the original 
Feynman integral.  For example, if in a five-point function two 
propagators $D_{1,2}$
never become singular, it is beneficial to 
combine them first into a single propagator: 
\be
\frac{1}{D_1 D_2} = \int \limits_{0}^{1} 
\frac{{\rm d}\xi}{D(\xi)^2},\;\;\;\
D(\xi) = \xi D_1 + (1-\xi) D_2.
\ee
Using this and similar techniques we are 
able to obtain a stable numerical representation for all virtual diagrams that 
contribute to the $gg \to t \bar t Z$ process.

In order to arrive at a finite result at next-to-leading order, 
renormalization is required.  We employ  the ${\overline {\rm MS}}$
renormalization constants for the QCD coupling $\alpha_s$ with five active 
flavors; the top quark contribution to the coupling constant renormalization 
is decoupled through a zero-momentum subtraction. In addition, we 
renormalize the top quark wave function and the top quark mass on shell.
No other renormalization constants or counter-terms are required to 
render ${\cal O}(\alpha_s)$ corrections to 
$gg \to t \bar t Z$ process ultraviolet finite.

We must also compute the real emission 
diagrams corresponding to the process $gg \to t \bar t Z + g$.  There are 50 such diagrams.  When the 
gluon in the final state is soft or collinear to the gluons in the initial 
state, the matrix element for $gg \to t \bar t Z + g$ process becomes 
singular.  To compute the corresponding contribution 
to the cross section, we split the phase-space for real gluon emission into 
soft, collinear and hard parts, 
using the two cutoff slicing method \cite{twocut}. The singularities that 
originate from collinear gluon emission are removed using 
the ${\overline {\rm MS}}$ renormalization of the parton distribution functions.
The computation of real emission corrections to $gg \to t \bar t Z$
using the two 
cutoff slicing method is analogous to  the computation of real emission 
corrections for the $gg \to t \bar t H$ process described in \cite{doreen}, 
where many details can be found.

\section{Results}

We now present the results of our computation.   
For all numerical results reported in this paper, 
we set the renormalization scale $\mu_r$ 
and the factorization scale $\mu_f$ equal to a common scale: $\mu_r = \mu_f = \mu$.
We employ the MRST parton distribution functions \cite{mrst} at the appropriate order in the perturbative expansion.  
We have applied a number of checks to our calculation.  
\begin{enumerate}
\item We have compared 
the leading order cross-section obtained with our code with 
the result of a similar computation using the program 
MadEvent~\cite{Maltoni:2002qb} and have found complete agreement.

\item The divergent parts of the NLO virtual correction to 
$ gg \to t\bar{t}Z$ can be expressed in a simple analytic form, as explained in 
detail in~\cite{doreen}.  We have checked that our results satisfy this check.

\item We have checked that all divergences 
cancel  once the real emission processes, 
the collinear counterterms, the renormalization constants, and the virtual corrections are combined.

\item An important check of our NLO virtual result is provided by its independence 
of the size of the contour deformation.

\item Finally, we have implemented 
all parts of the computation in several different codes that agree for all observables studied.

\end{enumerate}

\begin{tiny}
\begin{table}[htbp]
\vspace{0.1cm}
\begin{center}
\begin{tabular}{|c|c|c|c|}
\hline\hline
$\mu$ & $\sigma_{\rm LO}$, fb  & $\sigma_{\rm NLO}$, fb & 
$K = \sigma_{\rm NLO}/\sigma_{\rm LO}$  \\ \hline\hline
$\mu_0 / 8$ & 1046 & 587 & 0.56 \\ \hline
$\mu_0 / 4$ & 738 & 740 & 1.00 \\ \hline
$\mu_0 / 2$  & 537 & 739 & 1.38 \\ \hline
$\mu_0$    & 400 & 695 & 1.74 \\ \hline
$2\; \mu_0$    & 305 & 618 & 2.03 \\ \hline
\end{tabular}
\caption{\label{table1} The  cross section at leading and next-to-leading order 
for various values of the renormalization 
and factorization scales.  The numerical error on all numbers presented is $1\%$ or better.
}
\vspace{-0.1cm}
\end{center}
\end{table}
\end{tiny}

We begin with the total cross section. 
The results of our calculation are summarized in Table~\ref{table1}.
We show there the total cross section computed through 
leading and next-to-leading order for 
the five values of the scale $\mu = (1/8,1/4,1/2,1,2)\mu_0$,
where $\mu_0 = (2 m_t + m_Z)$.  For the QCD coupling constant, we use 
values consistent with the MRST parton densities at leading and next-to-leading order~\cite{mrst} at $\mu=m_Z$ and 
evolve these values using the beta-function at the appropriate order in the perturbative expansion. We see from Table~\ref{table1}
that the leading order result changes by more than a factor of $3$ 
when the scale $\mu$ changes between its minimal and maximal values.
This suggests significant ${\cal O}(\alpha_s)$ corrections.  Indeed, as clearly seen from the last column in Table~\ref{table1}, 
the NLO QCD corrections are large. They change the leading order 
corrections by up to a hundred percent, depending on the value 
of the scale $\mu$.  The dependence of the next-to-leading 
order cross section on the scale $\mu$ is significantly reduced; when 
$\mu$ is varied between $\mu_0/8$ and $2 \mu_0$, $\sigma_{\rm NLO}$
changes by only a factor of $1.25$. 

While formally all  values of the renormalization
and factorization scales are allowed, the best predictions for physical 
quantities at low orders in perturbation theory 
are obtained when the scales are set to  values 
suggested by  the physics of the problem.  The typical partonic 
center-of-mass  energies leading to the $t \bar t Z$ final state are moderate, 
$\sqrt{s_{gg}} \lesssim 10^3~{\rm GeV}$, and the typical transverse momenta
for the $Z$-boson and top quarks are $100-200~{\rm GeV}$.  It seems  reasonable to choose 
scales comparable to the typical transverse momenta.
This suggests setting $\mu \sim {\rm few} \times 100~{\rm GeV}$.  
Since $\mu_0 \sim 2m_t + m_Z \sim 450~{\rm GeV}$, we consider 
$\mu = [\mu_0/4,\mu_0]$ as a sensible range for the variation of the 
renormalization and factorization scales. 
As follows from Table~\ref{table1}, in this range $\sigma_{\rm LO}(\sigma_{\rm NLO})$ changes by about $\pm 30\%$($\pm 5\%$) 
if measured from the central value $\mu = \mu_0/2$.

\begin{figure}[htb]
\includegraphics[angle=90,width=3in,clip=]{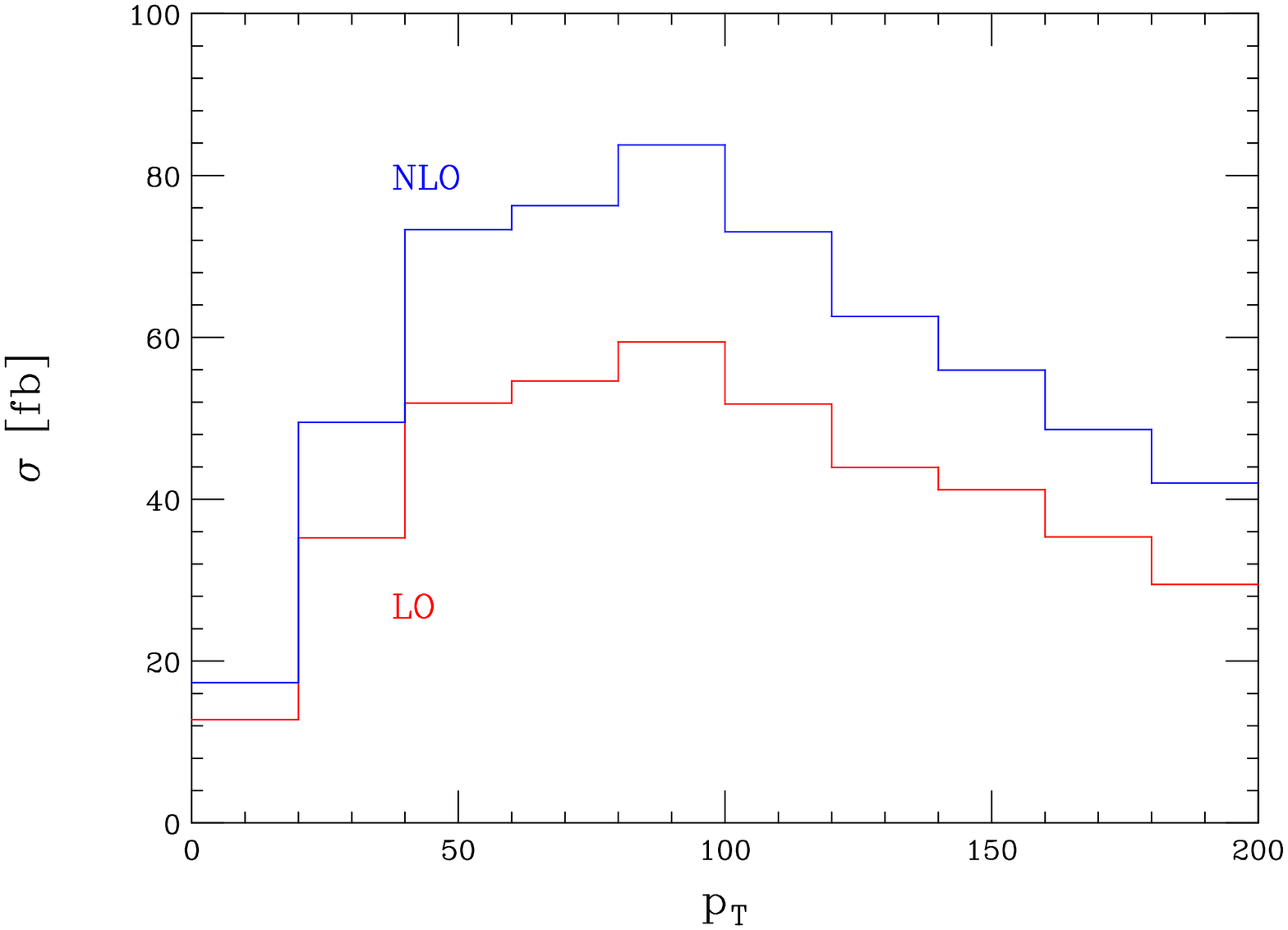}
\includegraphics[angle=90,width=3in,clip=]{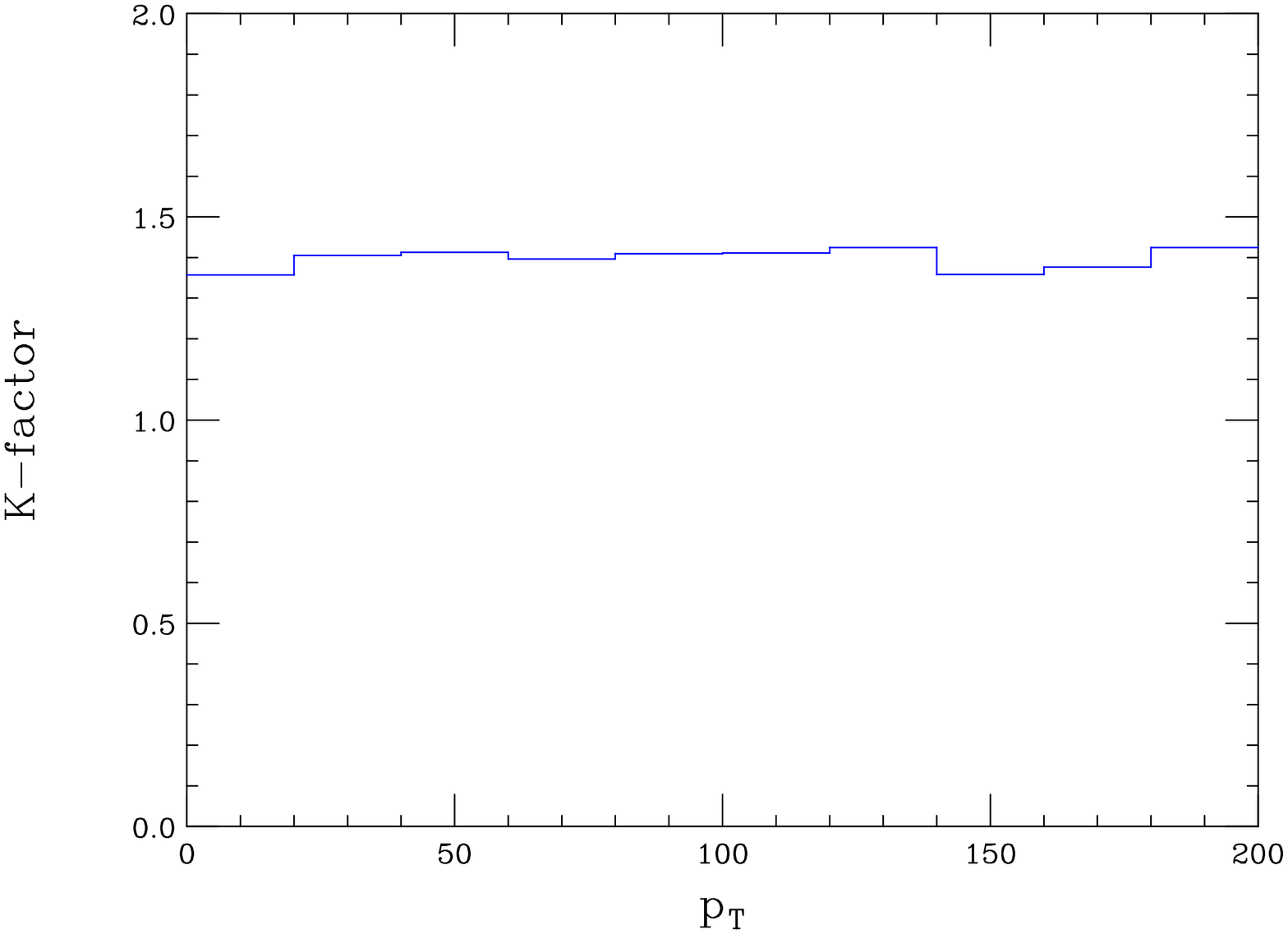}
\caption{  Left panel: the transverse momentum distribution of the $Z$-boson 
at leading order 
and next-to-leading order in $\alpha_s$.  Right panel: the dependence of the $K$-factor 
defined as $K = {\rm d}\sigma_{\rm NLO}/{\rm d}\sigma_{\rm LO}$ on the 
transverse momentum of the $Z$-boson. We set
$\mu = \mu_0/2$ in both plots.
} 
\end{figure}

As pointed out in Ref.~\cite{baur1,baur2}, the signal-to-background 
ratio for $pp \to t \bar t Z$ production is improved if a cut on the 
transverse momentum of the $Z$-boson $p_{\bot}(Z)$ 
is imposed since many background processes 
decrease faster than the signal when $p_{\bot}(Z)$ increases.  It is therefore interesting to  
study the effect of the NLO QCD corrections on the $p_{\bot}(Z)$ distribution.  In the left panel of 
Fig. 3 we show  the transverse momentum distribution of the $Z$-boson at 
leading order and next-to-leading order in the strong coupling constant.  We  combine 
events in $20~{\rm GeV}$ bins. We set the factorization 
and renormalization scales to $\mu = \mu_0/2$. While the QCD corrections 
change the distribution significantly, they affect only 
the overall normalization.  This is clearly seen from  the right 
panel of Fig. 3, where the $K$-factor defined as $K = {\rm d}\sigma_{NLO}/{\rm d}\sigma_{\rm LO}$ is shown.

We have only considered QCD corrections to the gluon fusion 
subprocess in $pp \to t \bar t Z$. Given the large gluon-gluon luminosity 
at the LHC and the fact that QCD corrections to gluon initiated 
processes  are usually larger than those for quark-initiated, this seems 
to be a reasonable first step towards the complete NLO QCD computation.  However, it is 
premature to discuss in detail how the NLO QCD corrections affect the analyses in Ref.\cite{baur1,baur2}.  
We make one comment here.  In Refs.\cite{baur1,baur2}, $\sigma(pp \to t \bar t Z)$ was computed  
with the renormalization and factorization scales set equal to the top quark mass.  
As follows from Table~\ref{table1}, for such renormalization and 
factorization scales
the NLO QCD corrections for the $gg \to t \bar t Z$ subprocess are
 about  $30$ percent. 
  Assuming that QCD corrections to 
quark intiated process are somewhat smaller, NLO QCD corrections of about $20\%$ to $pp \to t \bar t Z$ for $\mu=m_t$ can 
be expected. The NLO QCD corrections are positive and  
enhance the signal cross section, leading to a decrease in the luminosity 
required to probe the $t \bar t Z$ coupling.  The NLO QCD computation
also reduces the theoretical uncertainty of the signal cross section coming from uncalculated higher order corrections.  
We estimate that the residual scale uncertainty is $\pm 5\%$ by varying the scale between $\mu_0/4$ and $\mu_0$.

\section{Conclusions}
In this paper we presented the NLO QCD corrections 
to gluon-initiated production of $t\bar{t}Z$ at the LHC.  This is a first step towards the full NLO QCD 
calculation of $pp \to t\bar{t}Z$ needed to study electroweak top quark couplings.  
For reasonable choices of the renormalization and factorization scales, 
the QCD corrections can be up to $75\%$. The remaining 
theoretical uncertainty in the cross section after inclusion of the NLO corrections is estimated to be $\pm 5\%$.
The NLO QCD contributions are independent of the transverse momentum of the $Z$-boson.

In the future, we plan to extend  this computation to include 
the quark anti-quark annihilation and quark-gluon partonic scattering processes.  
It will then be possible 
to incorporate  the improved knowledge of the signal processes  into  the analysis of Ref.~\cite{baur1,baur2}. 

\vspace*{0.5cm}
{\bf Acknowledgments} 
K.M. would like to thank the Galileo Galilei Institute for Theoretical 
Physics for their hospitality, and the INFN for partial support during the completion of this work.
K.M. and A.L. are supported in part by the DOE grant DE-FG03-94ER-40833, Outstanding  Junior Investigator Award and by the Alfred P.~Sloan Foundation. 
F.P. is supported by the DOE grant DE-FG02-95ER40896, Outstanding  Junior Investigator Award, by the University of Wisconsin Research Committee
with funds provided by the Wisconsin Alumni Research Foundation, and
by the Alfred P.~Sloan Foundation.


\end{document}